\documentclass[12pt,aps,prd,showpacs,amsmath,amssymb]{revtex4}
\input epsf
\begin{document}
\title{Analysis of the semileptonic decay $\Lambda_c\rightarrow ne^+\nu_e$}
\author{Cheng-Fei Li$^1$, Yong-Lu Liu$^1$\footnote{corresponding author}, Ke Liu$^1$, Chun-Yu Cui$^2$, and Ming-Qiu Huang$^1$}
\affiliation{$^1$ College of Science, National University of Defense Technology, Hunan 410073, China}
\affiliation{$^2$ School of Biomedical Engineering, Third Military Medical University and Chongqing University, Chongqing 400038, China}
\date{\today}
\begin{abstract}
The semileptonic weak decay process of the $\Lambda_c$ baryon to the neutron $\Lambda_c\rightarrow ne^+\nu_e$ is examined. The transition form factors are investigated with light-cone QCD sum rules. The differential decay width is obtained in the dynamical region by fitting the sum rules-allowed results with the dipole formula. The total decay width and the branching ratio are estimated to be $\Gamma(\Lambda_c\rightarrow ne^+\nu_e)=(8.89\pm0.36)\times10^{-15}\,\mbox{GeV}$ and $\mbox{Br}(\Lambda_c\rightarrow ne^+\nu_e)=0.27\pm0.01\%$, respectively.
\end{abstract}
\pacs{11.25.Hf,~ 11.55.Hx,~ 13.40.Gp.} \maketitle

\section{Introduction}\label{sec1}

Heavy flavor physics may provide many details of the Standard Model, such as information of the Cabibbo-Kobayashi-Maskawa (CKM) matrix elements, the intrinsic structure of the strong interaction, possible evidences of new physics beyond the Standard Model, and so on. Investigation of heavy $b$ or $c$ mesons or baryons appeals much interest both in experiments and theory. There are many research works on the $\Lambda_c$ baryon in the past years. In experiments, data of the process $\Lambda_c\rightarrow pK^-\pi^+$ has been obtained and improved in fairly good accuracy. Early in 2013 the Belle Collaboration at KEKB measured the first model independent branching ratio $\mbox{Br}{(\Lambda_c\rightarrow pK^-\pi^+)}=(6.84\pm0.24^{+0.21}_{-0.27})\%$\cite{Belle}. This measurement improved the precision of the absolute branching ratios of other decay modes of the $\Lambda_c$ baryon. Next in 2015 the BESIII Collaboration reported the first absolute measurement of the branching ratio of the semileptonic decay process $\Lambda_c \rightarrow \Lambda e^+ \nu_e$, $\mbox{Br}{(\Lambda_c\rightarrow \Lambda e^+\nu_e)}=(3.63\pm0.38(sta)\pm0.20(sys))\%$\cite{Bes3}. Later in 2016, the BESIII Collaboration gives the improvement of the $\Lambda_c$ decay modes, including the process $\Lambda_c \rightarrow p K^-\pi^+$, $\mbox{Br}{(\Lambda_c\rightarrow pK^-\pi^+)}=(5.84\pm0.27\pm0.23)\%$\cite{Bes3-2}. Many theoretical analysis have been devoted to the topic, including the covariant quark model\cite{Gutsche,Gutsche2}, the light-cone QCD sum rules\cite{lcsr-lam,lcsr-azizi,Emmerich}, the SU(3) flavor symmety\cite{CDLV}, the relativistic quark model\cite{Faustov}, and other models\cite{Ikeno, Pervin}.

From the viewpoint of quark level, there are two main modes in heavy-light semiletonic decay channels at tree level of a baryon containg a $c$ quark, namely $c\rightarrow s$ and $c\rightarrow d$. Respectively, they correspond to the final baryons $\Lambda$ and $n$ in the case of the $\Lambda_c$ baryon decay modes. It is known that the decay amplitude is suppressed by the CKM matrix element $|V_{cq}|^2$, so the decay rate of the mode $\Lambda_c\rightarrow n$ is expected to be much smaller than that of $\Lambda_c\rightarrow \Lambda$ because $V_{cd}/V_{cs}\sim 0.2$. It may be one of the main reasons why this mode has not been measured experimentally yet to date. However, the lifetime of the neutron is much longer than that of the $\Lambda$ baryon, so it may be expected to test the process more accurately in experiments in spite of its small decay rate.

Furthermore, it can be expected that more data on the process $\Lambda_c\rightarrow n$ will appear in the near future, so it is useful to investigate the process theoretically in a reliable and well used method. Light-cone QCD sum rules has been well used to study the heavy-light semileptonic decay process. In Ref. \cite{lcsr-azizi} the author studied the form factors related to the process $\Lambda_Q\rightarrow Nl\nu_l$ with the light-cone QCD sum rules. The analysis shows that the results of the process $\Lambda_c\rightarrow ne^+\nu_e$ are larger than recent results from other approaches\cite{Gutsche,Pervin,CDLV,Faustov}. Therefore a reanalysis is useful and instructive for discussion of this process. In this paper we choose a different interpolating current for the $\Lambda_c$ baryon as inputs and expect to gain better estimate.

At the fundamental particle level, the process $\Lambda_c\rightarrow ne^+\nu_e$ can be divided into two parts. On the hadron side, we need to consider the weak decay process $c\rightarrow d$ via the $W^+$ boson, while on the lepton side the $W^+$ boson decays into $e^+\nu_e$. In calculations with effective theories, the leptonic part can be calculated with conventional field theory method, while the hadron part, which is generally parameterized using form factors, need to be analyzed taking into account the nonperturbative effects. Light-cone QCD sum rules is a useful nonperturbative method\cite{lcsr1,lcsr2,lcsr3,Colangelo}. When starting from the correlation function, which is the matrix element of the current-current interaction between vacuum and the final state, the decay quark fields are contracted, while the other ones between the vacuum and the final state can be expanded on the light-cone, which is related to the nonperturbative effect of the QCD. Generally the nonperturbative effect can be parameterized with some general functions which do not rely on any specific process. We call these functions light-cone distribution amplitudes(LCDAs). Light-cone distribution amplitudes denote the distribution of the hadron momentum among the partial particles. Therefore, the predictions with this method depends on the precision of the light-cone distribution amplitudes of the related particle.

The light-cone distribution amplitudes of the neutron has been examined for a long time. Since the beginning of this century, many efforts have been devoted to the study of the light-cone distribution amplitudes of the nucleon\cite{Braun,Lenz,dalattice} and the strange octet baryons\cite{DAs,LCDA-higherlam,Bali}. The reader is referred to the paper\cite{overviewofDAs} for a detailed overview. These LCDAS have been well used in analyzing the QCD related processes of the baryons\cite{LYL2,LYL3,WDW,Aliev,wym} and give fairly accurate predictions.

The rest of the manuscript is organized as follows. Section \ref{sec:sumrule} is devoted to the derivation of the sum rules of the weak decay form factors. In Sec. \ref{sec:analysis}, the sum rules are analyzed numerically and the main results are presented. A brief summary is presented in Sec. \ref{sec:summary}.

\section{Light-cone QCD sum rules of the process $\Lambda_c\rightarrow ne^+\nu_e$}\label{sec:sumrule}

The effective Hamitonian of the weak decay process $\Lambda_c\rightarrow ne^+\nu_e$ is
\begin{equation}
{\cal H}_{eff}=\frac{G_F}{\sqrt2}V_{cd}\bar d\gamma_\mu(1-\gamma_5)c\bar u_e\gamma^\mu(1-\gamma_5)\nu_e,
\end{equation}
where $G_F$ is the Fermi constant, $V_{cd}$ is the CKM matrix element, $d$, $c$, $u_e$ and $\nu_e$ are the field operators of the $d$, $c$, $e^+$ and $\nu_e$ particles. The main aim of this paper is to deal with the hadronic part which can be parameterized as the heavy-light quark transition form factors.

\begin{figure}
\begin{minipage}{8cm}
\epsfxsize=7cm \centerline{\epsffile{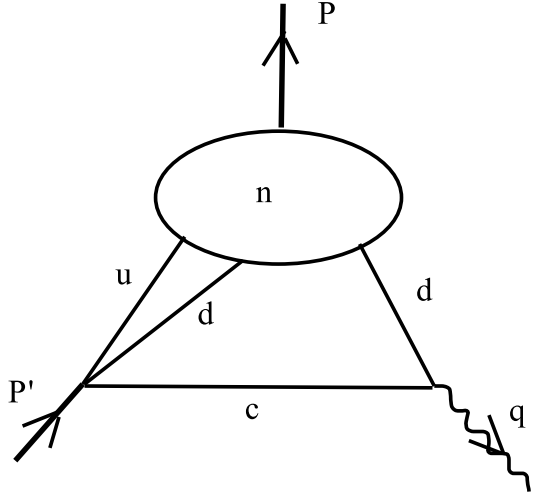}}
\end{minipage}
\caption{ Leading-order contribution to the correlation function (\ref{correlator}).} \label{figure-leading}
\end{figure}

In accordance to the standard procedure of the light-cone QCD sum rules, the derivation of the sum rules for the form factors starts from the following two-point correlation function between vacuum and the final state $|N\rangle$:
\begin{equation}
z^\mu T_\mu(P,q)=i z^\mu\int d^4xe^{iq\cdot x}\langle
0|T\{j_{\Lambda_c}(0)j_\mu(x)\}|N(P)\rangle,\label{correlator}
\end{equation}
where $j_{\Lambda_c}$ is the interpolating current of the $\Lambda_c$ baryon, $j_\mu(x)=\bar c\gamma_\mu(1-\gamma_5)d(x)$ is the current describing the weak interaction at the quark level, and $z$ is the light-cone vector with $z^2=0$. Fig. \ref{figure-leading} is a legend of the leading order contribution to the correlation function. In this paper we use the following current defined on the light-cone to interpolate the $\Lambda_c$ baryon field
\begin{equation}\label{intcurrent}
{j_{\Lambda_c}}(x)=\epsilon_{ijk}(u^i(x)C\gamma_5\!\not\!
zd^j(x))\!\not\! zc^k(x).
\end{equation}
The coupling constant of the baryon is defined by the matrix element of the
interpolating current (\ref{intcurrent}) between the vacuum and the baryon state:
\begin{equation}
\langle 0|j_{\Lambda_c}|\Lambda_c(P)\rangle=f_{\Lambda_c}(P\cdot z)\!\not\!
z\Lambda_c(P),\label{norm}
\end{equation}
where $f_{\Lambda_c}$
determines the normalization of the leading twist neutron
distribution amplitude.

Considering the quantum numbers of the baryons and the Lorentz structure of the current, the weak transition of the $\Lambda_c$ to the neutron can be parameterized as follows:
\begin{eqnarray} \label{ff}
\langle\Lambda_c(P-q)|j_\mu|n\rangle=\bar u_{\Lambda_c}(P-q)\Big[ f_1\gamma_\mu-i\frac{f_2}{M_{\Lambda_c}}\sigma_{\mu\nu}q^\nu-\frac{f_3}{M_{\Lambda_c}}q^\nu\nonumber\\
-\big(g_1\gamma_\mu+i\frac{g_2}{M_{\Lambda_c}}\sigma_{\mu\nu}q^\nu-\frac{g_3}{M_{\Lambda_c}}q^\nu\big)\gamma_5 \Big] N(P).
\end{eqnarray}
In the above expression, $f_i$ and $g_i$, $i=1,2,3$, are the weak transition form factors, $u_{\Lambda_c}$ and $N$ are the spinors of $\Lambda_c$ and the neutron, respectively, and $M_{\Lambda_c}$  is the mass of $\Lambda_c$. At hadronic level, the correlation function (\ref{correlator}) is calculated by inserting a complete set of baryon states which have the same quantum numbers as $\Lambda_c$. By using the translation invariance and integrating the coordinate variables, the hadronic part is obtained as
\begin{eqnarray}
z_\mu T^\mu(P,q)=\frac{2f_{\Lambda_c}}{M_{\Lambda_c}^2-P'^2}(P'\cdot z)^2\Big[f_1\!\not\!z+f_2\frac{\!\not\! z\!\not\! q}{M_{\Lambda_c}}-\big(g_1\!\not\! z-g_2\frac{\!\not\! z\!\not\!q}{M_{\Lambda_c}}\big)\gamma_5\Big]N(P)+\cdots,
\end{eqnarray}
in which $``\cdots"$ denotes the resonance and continuum contributions and $P'=P+q$. The form factors $f_3$ and $g_3$ vanish in the representation due to the conservation of the vector current.

On the other hand, the correlation function (\ref{correlator}) can be expanded on the light-cone $z^2=0$. With the light-cone distribution amplitudes presented in Refs.\cite{Braun,Lenz}, the correlation function (\ref{correlator}) is calculated at the QCD level as
\begin{eqnarray}
z_\mu T^\mu(P,q)&=&\int_0^1d\alpha_2\frac{\alpha_2}{(\alpha_2P-q)^2-m_c^2}\Big\{B_0(\alpha_2)-\frac{\alpha_2M^2}{(\alpha_2P-q)^2
-m_c^2}B_1(\alpha_2)\nonumber\\
&&+\frac{M^2}{(\alpha_2P-q)^2-m_c^2}B_3(\alpha_2)+\frac{2\alpha_2^2M^4}{((\alpha_2P-q)^2-m_c^2)^2}B_4(\alpha_2)\nonumber\\
&&-\frac{2\alpha_2m_c^2M^2}{((\alpha_2P-q)^2-m_c^2)^2}B_3(\alpha_2)\Big\}(P\cdot z)^2(\!\not\! z(1-\gamma_5)N)\nonumber\\
&&+\int_0^1d\alpha_2\frac{\alpha_2M}{(\alpha_2P-q)^2-m_c^2}\Big\{B_2(\alpha_2)-\frac{2\alpha_2M^2}{((\alpha_2P-q)^2-m_c^2)^2}B_4(\alpha_2)\Big\}\nonumber\\
&&\times(P\cdot z)^2(\!\not\! z\!\not\! q(1+\gamma_5)N),\label{lc-expansion}
\end{eqnarray}
where $M$ is the mass of the neutron and the light quark is approximated to be massless $m_{u,d}=0$, and the following abbreviates are used for compactness:
\begin{eqnarray}
B_0(\alpha_2)&=&\int_0^{1-\alpha_2}d\alpha_1(V_1+A_1+2T_1)(\alpha_1,\alpha_2,1-\alpha_1-\alpha_2),\nonumber\\
B_1(\alpha_2)&=&2\widetilde V_1(\alpha_2)-\widetilde
V_2(\alpha_2)-\widetilde V_3(\alpha_2)-\widetilde
V_4(\alpha_2)-\widetilde
V_5(\alpha_2)+2\widetilde A_1(\alpha_2)-\widetilde
A_2(\alpha_2)+\widetilde A_3(\alpha_2)\nonumber\\
&&+\widetilde
A_4(\alpha_2)-\widetilde
A_5(\alpha_2)+2\widetilde T_1(\alpha_2)+\widetilde
T_2(\alpha_2)-3\widetilde T_3(\alpha_2)-2\widetilde
T_4(\alpha_2)-\widetilde
T_5(\alpha_2)+2T_7(\alpha_2),\nonumber\\
B_2(\alpha_2)&=& \widetilde V_1(\alpha_2)-\widetilde
V_2(\alpha_2)-\widetilde V_3(\alpha_2)+\widetilde A_1(\alpha_2)-\widetilde
A_2(\alpha_2)+\widetilde A_3(\alpha_2)+\frac32\widetilde T_1(\alpha_2)+\frac12\widetilde
T_2(\alpha_2)\nonumber\\
&&-2\widetilde T_3(\alpha_2)-\widetilde
T_5(\alpha_2),\nonumber\\
B_3(\alpha_2)&=&2\widetilde{\widetilde
T}_2(\alpha_2)-3\widetilde{\widetilde
T}_3(\alpha_2)-2\widetilde{\widetilde
T}_4(\alpha_2)+2\widetilde{\widetilde T}_5(\alpha_2)
+2\widetilde{\widetilde T}_7(\alpha_2)+2\widetilde{\widetilde
T}_8(\alpha_2),\nonumber\\
B_4(\alpha_2)&=&\widetilde{\widetilde
V}_1(\alpha_2)-\widetilde{\widetilde
V}_2(\alpha_2)-\widetilde{\widetilde
V}_3(\alpha_2)-\widetilde{\widetilde
V}_4(\alpha_2)-\widetilde{\widetilde V}_5(\alpha_2)
+\widetilde{\widetilde V}_6(\alpha_2)+{\widetilde
A}_1(\alpha_2)-\widetilde{\widetilde
A}_2(\alpha_2)\nonumber\\
&&+\widetilde{\widetilde A}_3(\alpha_2)
+\widetilde{\widetilde A}_4(\alpha_2)-\widetilde{\widetilde
A}_5(\alpha_2)+\widetilde{\widetilde
A}_6(\alpha_2)-2\widetilde{\widetilde
T}_1(\alpha_2)+2\widetilde{\widetilde
T}_2(\alpha_2)-3\widetilde{\widetilde T}_3(\alpha_2)\nonumber\\
&&-2\widetilde{\widetilde T}_4(\alpha_2)+4\widetilde{\widetilde
T}_5(\alpha_2)-2\widetilde{\widetilde
T}_6(\alpha_2)+6\widetilde{\widetilde
T}_7(\alpha_2)+6\widetilde{\widetilde
T}_8(\alpha_2),
\end{eqnarray}
where the definitions of the distribution amplitudes $V_i, A_i, T_i$ with definite twist can be found in Refs.\cite{Braun,Lenz}, and the functions with a ``tilde" are defined as
\begin{eqnarray}
\widetilde
F(\alpha_2)&=&\int_0^{\alpha_2}d\alpha_2'\int_0^{1-\alpha_2'}d\alpha_1F(\alpha_1,\alpha_2',1-\alpha_1-\alpha_2'),\nonumber\\
\widetilde{\widetilde
F}(\alpha_2)&=&\int_0^{\alpha_2}d\alpha_2'\int_0^{\alpha_2'}d\alpha_2''\int_0^{1-\alpha_2''}d\alpha_1 F(\alpha_1,\alpha_2'',1-\alpha_1-\alpha_2'').
\end{eqnarray}

To get reliable and stable sum rules, quark-hadron duality and Borel transformation are usually adopted. The main idea of the duality approximation is that the spectral density of the hadron part can be approximated by the spectral density calculated in QCD, so that the higher resonance and continuum contributions in hadron representation can be calculated with the integration of the spectral density upon some value that is related to the first resonance of the intermediate hadron we insert, and the threshold $s_0$ is thus introduced to denote this value. Another parameter $M_B^2$ comes from the Borel transformation, which is defined as
\begin{equation}
\hat{B}^{(Q^2)}_{M_B^2}\equiv \lim_{Q^2\rightarrow \infty,N\rightarrow \infty}\frac{1}{\Gamma(N)}(-Q^2)^N(\frac{d}{dQ^2})^N,
\end{equation}
where $Q^2$ is the variable and $M_B^2=Q^2/N$. With the Borel transformation, both the higher dimension or higher twist and higher resonance contributions are suppressed at a reasonable working window of the Borel mass $M_B^2$. Considering the quark-hadron duality and the Borel transformation, the two representations are matched together to get the sum rules of the form factors. In practice the following formulae are used:
\begin{eqnarray}\label{formula}
\int_0^1 dx\frac{\rho(x)}{(q-xP)^2-m^2}&=&-\int_0^1\frac{dx}{x}\frac{\rho(x)}{(s'-P'^2)}\rightarrow-\int_{x_0}^1\frac{dx}{x}\rho(x)e^{-s'/M_B^2},\nonumber\\
\int_0^1 dx\frac{\rho(x)}{((q-xP)^2-m^2)^2}&=&\int_0^1\frac{dx}{x^2}\frac{\rho(x)}{(s'-P'^2)^2}\rightarrow\frac{1}{M_B^2}\int \frac{dx}{x^2}\rho(x)e^{-s'/M_B^2}\nonumber\\
&&+\frac{\rho(x_0)e^{-s_0/M_B^2}}{x_0^2M^2-q^2+m^2},\nonumber\\
\int_0^1 dx\frac{\rho(x)}{((q-xP)^2-m^2)^3}&=&-\int_0^1\frac{dx}{x^3}\frac{\rho(x)}{(s'-P'^2)^3}\rightarrow-\frac{1}{2M_B^4}\int_{x_0}^1\frac{dx}{x^3}\rho(x)e^{-s'/M_B^2}\nonumber\\
&&-\frac12\frac{\rho(x_0)e^{-s_0/M^2}}{x_0M_B^2(x_0^2M^2-q^2+m^2)}\nonumber\\
&&+\frac12\frac{x_0^2 e^{-s_0/M_B^2}}{x_0^2M^2-q^2+m^2}\left[\frac{d}{dx}\frac{\rho(x_0)}{x_0(x_0^2M^2-q^2+m^2)}\right],\nonumber\\
\end{eqnarray}
in which $s'=(1-x)M^2-\frac{1-x}{x}q^2+\frac{m^2}{x}$, $x_0=\frac{-(-q^2+s_0-M^2)+\sqrt{(-q^2+s_0-M^2)^2+4M^2(-q^2+m^2)}}{2M^2}$.

The final sum rules for the form factors are
\begin{eqnarray}\label{sumrule1}
2f_{\Lambda_c}f_1e^{-\frac{M_{\Lambda_c}^2}{M_B^2}}&=&\int_{\alpha_{20}}^1d\alpha_2e^{\frac{s'}{M_B^2}}\Big\{-B_0(\alpha_2)-\frac{M^2}{M_B^2}B_1(\alpha_2)+\frac{M^2}{\alpha_2M_B^2}B_3(\alpha_2)\nonumber\\
&&-\frac{M^4}{M_B^4}B_4(\alpha_2)+\frac{m_c^2M^2}{\alpha_2^2M_B^4}B_3(\alpha_2)\Big\}+\frac{Me^{s_0/M_B^2}}{\alpha_{20}^2M^2-q^2+m_c^2}\nonumber\\
&&\times\Big\{-\alpha_{20}^2B_1(\alpha_{20})+\alpha_{20}B_3(\alpha_{20})-\frac{\alpha_{20}^2M^2}{M_B^2}B_4(\alpha_{20})+\frac{m_c^2}{M_B^2}B_3(\alpha_{20})\Big\}\nonumber\\
&&+\frac{\alpha_{20}^2M^2e^{s_0/M_B^2}}{\alpha_{20}^2M^2-q^2+m_c^2}\frac{d}{d\alpha_{20}}\Big\{\frac{\alpha_{20}^2M^2B_4(\alpha_{20})-m_c^2B_3(\alpha_{20})}{\alpha_{20}^2M^2-q^2+m_c^2}\Big\}
\end{eqnarray}
and
\begin{eqnarray}\label{sumrule2}
2f_{\Lambda_c}f_2e^{-\frac{M_{\Lambda_c}^2}{M_B^2}}&=&M_{\Lambda_c} \int_{\alpha_{20}}^1d\alpha_2\frac{e^{\frac{s'}{M_B^2}}}{\alpha_2}\Big\{\frac{M}{M_B^2}B_0(\alpha_2)+\frac{M^3}{M_B^4}B_4(\alpha_2)\Big\}\nonumber\\
&&+\frac{\alpha_{20}Me^{s_0/M_B^2}}{\alpha_{20}^2M^2-q^2+m_c^2}\Big\{B_2(\alpha_{20})+\frac{M^2}{M_B^2}B_4(\alpha_{20})\Big\}\nonumber\\
&&-\frac{\alpha_{20}^2M^3e^{s_0/M_B^2}}{\alpha_{20}^2M^2-q^2+m_c^2}\frac{d}{d\alpha_{20}}\Big\{\frac{\alpha_{20}B_4(\alpha_{20})-m_c^2B_3(\alpha_{20})}{\alpha_{20}^2M^2-q^2+m_c^2}\Big\}.
\end{eqnarray}
In the expressions $\alpha_{20}$ is the same as that of $x_0$ in Eq. (\ref{formula}). It is also noted that calculations show that the axial form factors $g_i(q^2)$ satisfy the relation $g_i(q^2)=f_i(q^2)$ for $i=1,2$.

\section{Numerical analysis of the sum rules}\label{sec:analysis}

The most important inputs in the light-cone QCD sum rules are the light-cone distribution amplitudes of the hadron. In the following analysis, we use definitions and the nonperturbative parameters presented in Refs. \cite{Braun, Lenz, WDW}. We do not show them explicitly in this paper, and those interested in the contents are referred to the references above.

\begin{figure}
\begin{minipage}{8cm}
\epsfxsize=7cm \centerline{\epsffile{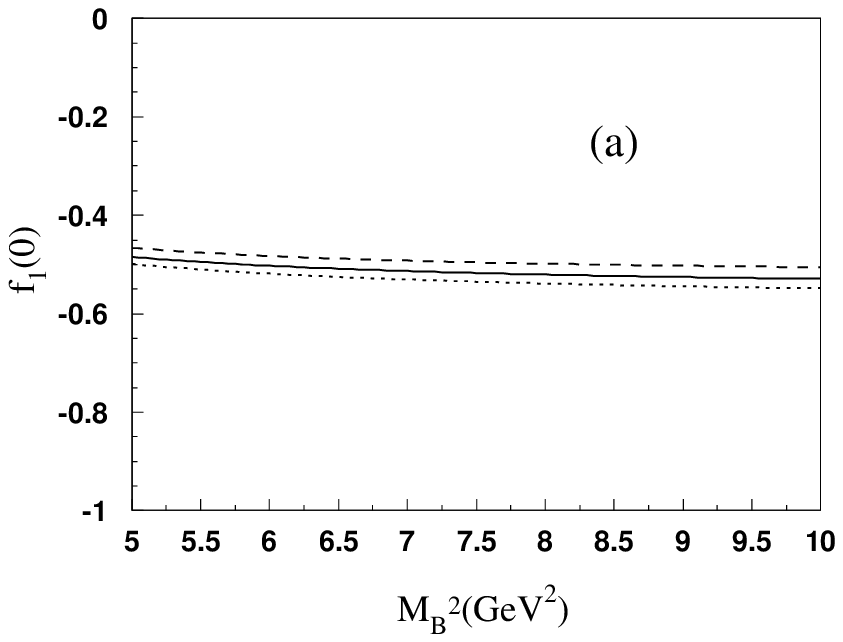}}
\end{minipage}
\begin{minipage}{8cm}
\epsfxsize=7cm \centerline{\epsffile{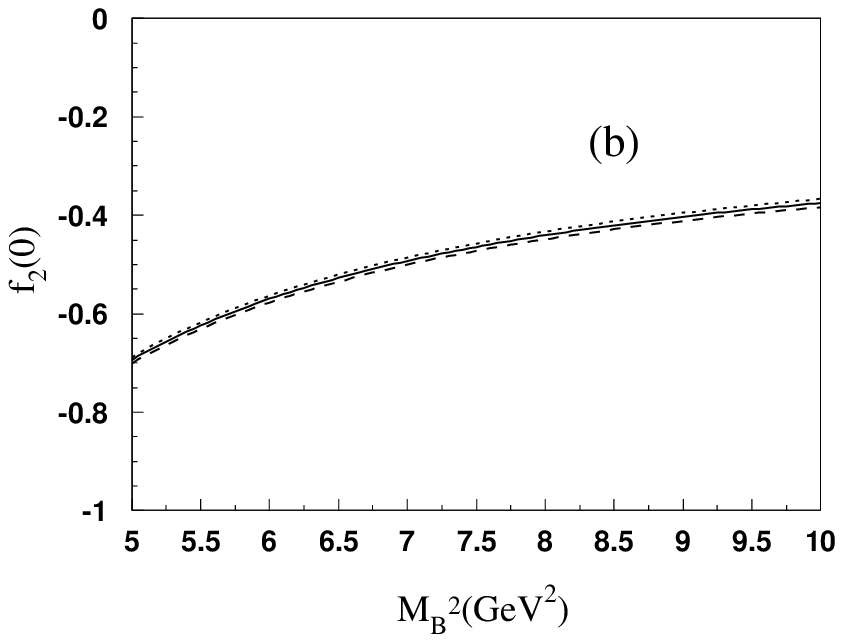}}
\end{minipage}
\caption{ Dependence of the form factors $f_1(0)$ (a) and $f_2(0)$ (b) on the Borel parameter. The dashed, solid, and dotted lines correspond to
the points $s_0^2=7.5,\;8,\;8.5\;\mbox{GeV}^{2}$, respectively.} \label{fig:1}
\end{figure}

\begin{figure}
\begin{minipage}{8cm}
\epsfxsize=7cm \centerline{\epsffile{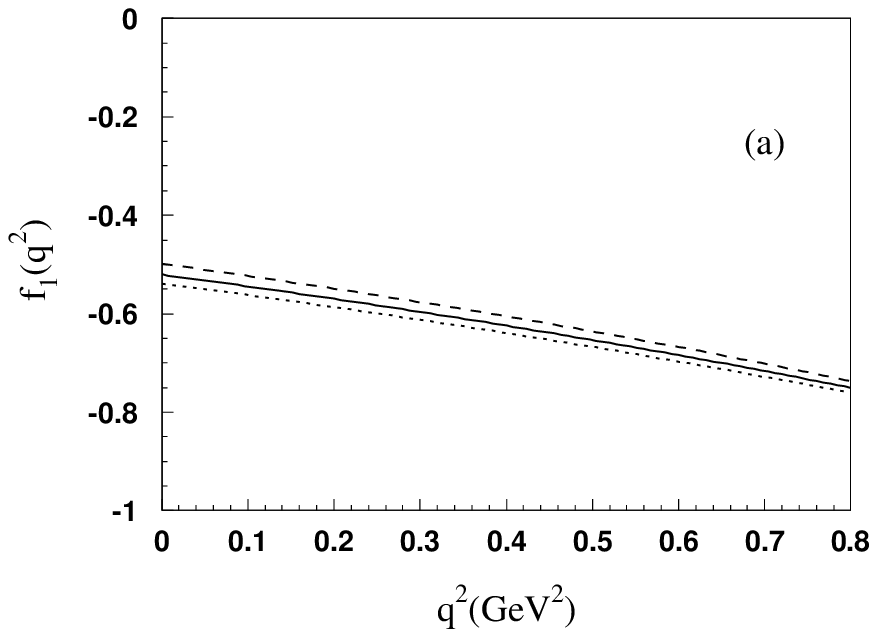}}
\end{minipage}
\begin{minipage}{8cm}
\epsfxsize=7cm \centerline{\epsffile{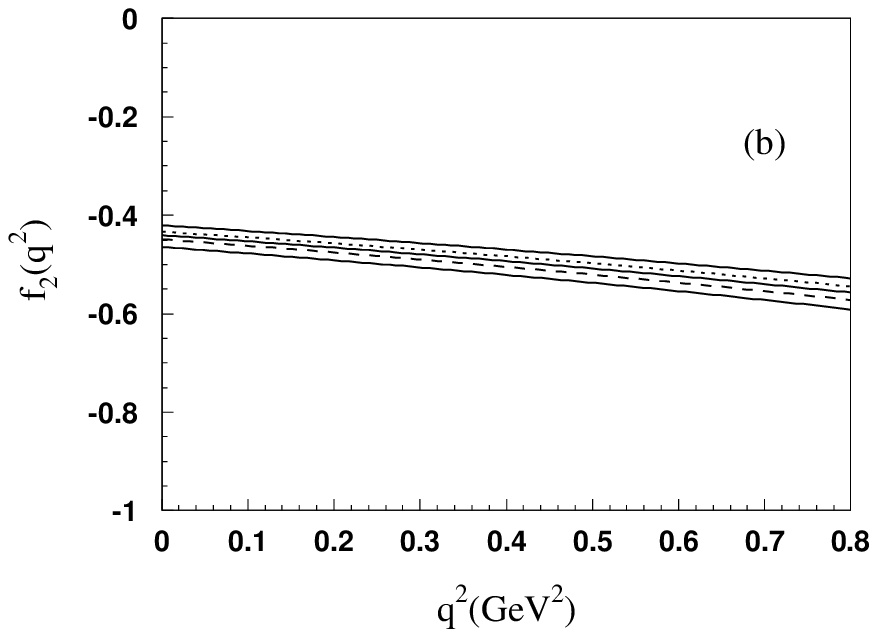}}
\end{minipage}
\caption{(a) $q^2$ dependence of $f_1(q^2)$ with $M_B^2=8\,\mbox{GeV}^2$. The dashed, solid, and dotted lines correspond to
the points $s_0^2=7.5,\;8,\;8.5\;\mbox{GeV}^{2}$. (b) $q^2$ dependence of $f_2(q^2)$. The middle three lines are the same as that of (a). The upper and lower solid line represents $M_B^2=7.5\;,8.5\,\mbox{GeV}^2$ with $s_0=8\,\mbox{GeV}^2$.} \label{fig:2}
\end{figure}

There are still some other parameters which need to be determined. In the numerical analysis, we adopt the masses of the charm quark, the neutron and the $\Lambda_c$ baryon as $m_c=1.27\,\mbox{GeV}$, $M=m_n=0.939\,\mbox{GeV}$ and $M_{\Lambda_c}=2.286\;\mbox{GeV}$. The coupling constant $f_{\Lambda_c}$ defined in Eq. (\ref{norm}) is another important parameter to be determined. In the following analysis we use the estimate obtained in Ref. \cite{lcsr-lam}: $f_{\Lambda_c}=(9.1\pm0.2)\times10^{-3}\;\mbox{GeV}^2$. In addition, two artificial parameters, the continuous threshold $s_0$ and the Borel mass $M_B^2$, need to be determined. The threshold $s_0$ is the parameter to be used to represent the continuous contribution by integrating the spectral density in the area above it. In the viewpoint of the physical content, the threshold is related to the first excited state of the inserted ground hadron, the $\Lambda_c$ baryon. Additionally, the result should not depend on the threshold two much. In the work, we choose the value in the region $7.5\,\mbox{GeV}^2\le s_0\le 8.5\,\mbox{GeV}^2$. Another parameter, the Borel mass, is determined with the requirements that both higher twist contributions and higher resonance contributions are suppressed, and the sum rules do not depend on the Borel mass too much in the working region simultaneously. Making use of the light-cone distribution amplitudes presented in Ref.\cite{Braun, Lenz, WDW}, we show the dependence of the form factors on the Borel parameter at the squared momentum transfer $q^2=0\,\mbox{GeV}^2$ in Fig. \ref{fig:1}. The results indicate that in this window the sum rules of the form factor $f_1(0)$ meets the above requirements and are flat on the Borel parameter. However, the form factor $f_2(0)$ depends much on the Borel mass in the same region. The working window of Borel mass is thus chosen as $7.5\,\mbox{GeV}^2\le M_B^2\le 8.5\,\mbox{GeV}^2$. To make a more reliable estimate of the decay mode, we plot the $q^2$ dependence of the form factors $f_1(q^2)$ with the Borel mass being chosen as $M_B^2=8.0\,\mbox{GeV}^2$ and $f_2(q^2)$ with the three different points of Borel mass $M_B^2=7.5\,,8\,,8.5\,\mbox{GeV}^2$ in Fig. \ref{fig:2}.

When calculating the correlation function at the quark level, we need to expand it on the light-cone with twists. The higher twist contribution is suppressed by a factor $\frac{1}{m_c^2-q^2}$ compared with the lower twist one. Therefore the expansion is only reliable in the region where $m_c^2-q^2$ is large enough. In other words, the sum rules can only be well used in some range of the momentum transfer. In the analysis we choose the region of $q^2\le0.8 \mbox{GeV}^2$ under the requirement that $m_c^2-q^2>m_c\Lambda_{QCD}$. In order to get the form factors in the whole dynamical area, we need to extrapolate the results with some methods. The basic idea of the extrapolation is that the form factors are assumed to be smooth in the dynamic region so that the analytical results obtained in a relatively narrow region can be fitted with some polynomial functions. We firstly fit the sum rules in the region $0\le q^2\le 0.8 \mbox{GeV}^2$ with a general dipole formula
\begin{equation}\label{fitformula1}
f_i(q^2)=\frac{f_i(0)}{1+a_1q^2/M_{\Lambda_c}^2+a_2(q^2/M_{\Lambda_c}^2)^2}.
\end{equation}

In order to get reliable results, the fitting coefficients are determined as follows. For given Borel mass $M_B^2$ and threshold $s_0$, the fitting formula (\ref{fitformula1}) is used to give a set of values. With different choice of $M_B^2$ and $s_0$, several estimations of the coefficients are obtained. The central values of the coefficients come from the average of different fitting results. The errors come from the uncertainties of the inputs $M_B^2$ and $s_0$. Tab. \ref{fit1} presents the fitting results with the parameters $7.5\,\mbox{GeV}^2\leq M_B^2\leq 8.5\,\mbox{GeV}^2$ and $7.5\,\mbox{MeV}^2\le s_0\le 8.5\,\mbox{GeV}^2$. With the fitted coefficients, the form factors can be extrapolated to the whole dynamic region.

\begin{table}
\caption{ The fit for $f_1(q^2)$ and $f_2(q^2)$ with formula (\ref{fitformula1}).}\label{fit1}
\begin{tabular}{|c|c|c|c|}
  \hline
       &  $f_i(0)$   &   $a_1$ & $a_2$ \\
  \hline
  $f_1$  &  $-0.518\pm0.020$  &      $-2.372\pm0.152$     &       $2.536\pm0.355$ \\
  \hline
 $f_2$ &  $-0.441\pm0.008$  &      $-1.428\pm0.055$      &       $0.286\pm0.081$ \\
  \hline
\end{tabular}
\end{table}

The sum rules can also be fitted with some other formulae. In fact, the fitting formula (\ref{fitformula1}) is a general form of the dipolar one\cite{Becirevic}
\begin{equation}\label{fitformula2}
f_i(q^2)=\frac{f_i(0)}{(1-q^2/M_{\Lambda_c}^2)(1-\alpha q^2/M_{\Lambda_c}^2)}.
\end{equation}
With the same method for the central values and the errors of the fitting coefficients, we got the results in Tab. \ref{fit2}.
\begin{table}
\caption{ The fit for $f_1(q^2)$ and $f_2(q^2)$ with formula (\ref{fitformula2}).}\label{fit2}
\begin{tabular}{|c|c|c|}
  \hline
       &  $f_i(0)$   &   $\alpha$ \\
  \hline
  $f_1$  &  $-0.521\pm0.020$  &      $1.187\pm0.120$     \\
  \hline
 $f_2$ &  $-0.441\pm0.008$  &      $0.443\pm0.041$      \\
  \hline
\end{tabular}
\end{table}

\begin{figure}
\begin{minipage}{8cm}
\epsfxsize=7cm \centerline{\epsffile{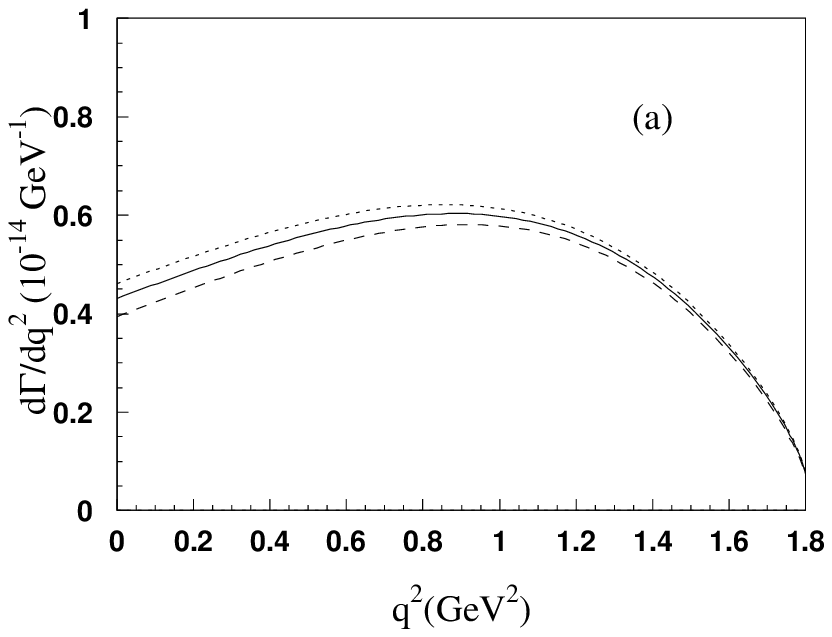}}
\end{minipage}
\begin{minipage}{8cm}
\epsfxsize=7cm \centerline{\epsffile{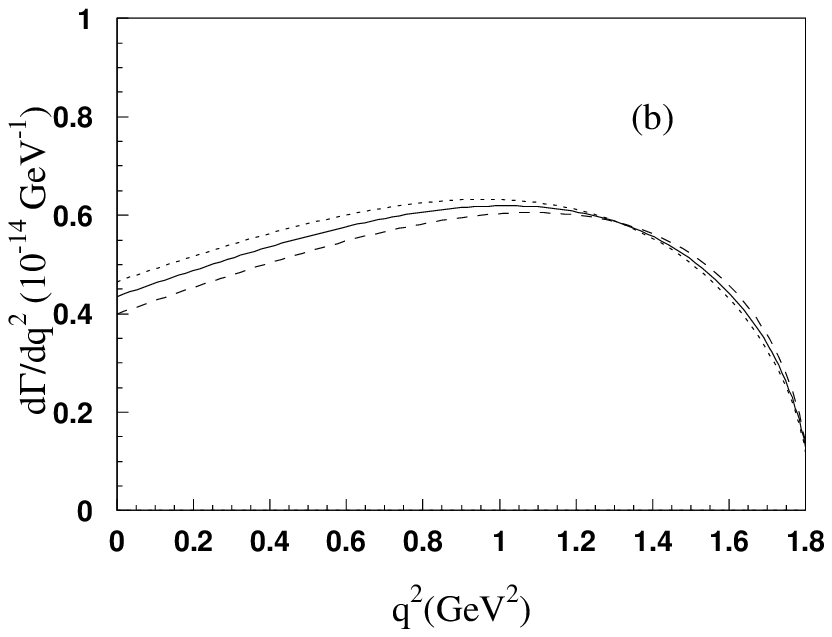}}
\end{minipage}
\caption{Differential decay width of the process $\Lambda_c\rightarrow ne^+\nu_e$. (a) illustrates results from Tab. \ref{fit1} and (b) is that of Tab. \ref{fit2}} \label{fig:3}
\end{figure}

The differential semileptonic decay width of the process is determined with the weak transition form factors by the following formula\cite{LYL2,WDW}
\begin{eqnarray}
\frac{d\Gamma}{dq^2}&=&\frac{G_F^2|V_{cd}|^2}{192\pi^3M_{\Lambda_c}^5}q^2\sqrt{q_+q_-}\Big\{-6f_1f_2M_{\Lambda_c}m_+q_-^2+6g_1g_2M_{\Lambda_c}m_-q_+^2
+f_1^2M_{\Lambda_c}^2\Big(\frac{m_+^2m_-^2}{q^2}+m_-^2\nonumber\\
&&-2(q^2+2M_{\Lambda_c}M_\Lambda)\Big)+g_1^2M_{\Lambda_c}^2\Big(\frac{m_+^2m_-^2}{q^2}+m_-^2-2(q^2-2M_{\Lambda_c}M_\Lambda)\Big)-f_2^2\Big(-2m_+^2m_-^2\nonumber\\
&&+m_+^2q^2+q^2(q^2+4M_{\Lambda_c}M_\Lambda)\Big)-g_2^2\Big(-2m_+^2m_-^2+m_-^2q^2+q^2(q^2-4M_{\Lambda_c}M_\Lambda)\Big)\Big\},\label{ddecay}
\end{eqnarray}
where $m_\pm=(M_{\Lambda_c}\pm M)$ and $q_\pm=(q^2-m_\pm^2)$ are used for convenience. In analysis, the parameters appearing in the formula are adopted the central values presented in PDG\cite{PDG}: the Fermi constant $G_F=1.1663787\times10^{-5}\,\mbox{GeV}^{-2}$, the CKM matrix element $V_{cd}=0.225$. With the fitting parameters presented in Tab. \ref{fit1} and Tab. \ref{fit2}, we plot the differential decay width of the process on the momentum transfer $q^2$ in Fig. \ref{fig:3}.

After integrating the differential decay width on the whole dynamic region $0\le q^2\le (M_{\Lambda_c}-M)^2$, we obtain the prediction of the total decay width $\Gamma(\Lambda_c\rightarrow ne^+\nu_e)=(8.89\pm0.36)\times10^{-15}\,\mbox{GeV}$ with fit (\ref{fitformula1}) and $\Gamma(\Lambda_c\rightarrow ne^+\nu_e)=(9.52\pm0.24)\times10^{-15}\,\mbox{GeV}$ with fit (\ref{fitformula2}), both of which are much less than that of the mode $\Lambda_c\rightarrow\Lambda e^+\nu_e$ \cite{Bes3}. As the fit formula (\ref{fitformula1}) is a general form of the dipole formula, it is reasonable to assume its prediction is more reliable. Furthermore, by using the mean life time of $\Lambda_c$ presented in PDG \cite{PDG} $\tau=(200\pm6)\times10^{-15}\mbox s$, we estimate the branching ratio of the process $\mbox{Br}(\Lambda_c\rightarrow ne^+\nu_e)=0.27\pm0.01\%$ with results from Tab. \ref{fit1}. It is shown in Tab. \ref{tab:3} the results from some other models. It can be seen that our estimates are consistent with other predictions and the results are steady with the two fit formulae. Although the branching ratio of this process is very small, it is still important because the long lifetime of the neutron makes it possible to measure the result more accurately than other channels.
\begin{table}
\caption{ Branching ratio of the process $\Lambda_c\rightarrow ne^+\nu_e$ from different references.}\label{tab:3}
\begin{tabular}{|c|c|c|c|c|c|}
  \hline
    Model   & \cite{Gutsche}   &  \cite{Pervin} &  \cite{CDLV} & \cite{Faustov}&  This work\\
  \hline
 $\mbox{Br}(\%)$ &  $0.236$  &      $0.27$      &       $0.293$ & $0.268$  &    $0.27\pm0.01$\\
  \hline
\end{tabular}
\end{table}

\section{Summary}\label{sec:summary}

The semileptonic decay of the $\Lambda_c$ baryon is a useful channel to test the Standard Model and the intrinsic structure of the strong interaction. Considering the recent development in experiments on the decay modes of $\Lambda_c$, we investigate the weak transition form factors of the channel $\Lambda_c\rightarrow ne^+\nu_e$ in the framework of light-cone QCD sum rules. We present the dependence of the form factors on the momentum transfer in the region where the sum rules are valid. Furthermore, we fit the form factors with a dipole formula and extrapolate them into the whole dynamic region. The differential decay width is presented in the whole dynamic region with the fitted parameters. The total decay width is obtained by integrating the differential decay width in the dynamic range $0\le q^2\le (M_{\Lambda_c}-M)^2$ to be $\Gamma(\Lambda_c\rightarrow ne^+\nu_e)=(8.89\pm0.36)\times10^{-15}\,\mbox{GeV}$ and the branching ratio is estimate with the aid of the mean life time of the $\Lambda_c$ baryon to be $\mbox{Br}(\Lambda_c\rightarrow ne^+\nu_e)=0.27\pm0.01\%$. The results show that this decay mode is fairly large and deserves to be measured in experiments.

\acknowledgments  This work was supported in part by the National
Natural Science Foundation of China under Contract Nos.11475257, 11675263, 11547179, and the National University of Defense Technology Foundation under Grant No. JC14-02-05.

\end{document}